\documentclass[prl,twocolumn,floatfix,superscriptaddress,showpacs,amsmath,amssymb]{revtex4}
\usepackage{bm}
\usepackage{graphicx}
\usepackage{dcolumn}
\usepackage{amsmath}
\usepackage{amssymb}
\usepackage{epsfig}

\begin{document}
\title{Spreading of a granular droplet}
\author{Iv\'{a}n S\'{a}nchez}
\affiliation{Departamento de F\'{\i}sica, Universidad Sim\'{o}n Bol\'{\i}var, Apartado 89000, Caracas 1080-A, Venezuela}
\author{Franck Raynaud}
\altaffiliation[presently at]{ MSC, UMR 7057 (CNRS), Univ.~Paris 7}
\author{Jos\'{e} Lanuza}
\author{Bruno Andreotti}
\author{Eric Cl\'{e}ment}
\affiliation{PMMH, UMR7636 (CNRS), ESPCI Univ.~P6-P7, 10 Rue Vauquelin, 75005 Paris, France. }
\author{Igor S. Aranson}
\affiliation{Materials Science Division, Argonne National Laboratory, Argonne, IL60439, USA}

\date{\today}

\begin{abstract}
The influence of controlled vibrations on the granular rheology is investigated in a specifically designed experiment. We study experimentally a thin granular film spreading under the action of horizontal vibrations. A nonlinear diffusion equation is derived theoretically that describes the evolution of the deposit shape. A self-similar parabolic shape (the``granular droplet'') and a spreading dynamics are predicted that both agree quantitatively with the experimental results. The theoretical analysis is used to extract effective dynamic friction coefficients between the base and the granular layer under sustained and controlled vibrations. We derive an empirical friction law involving external driving parameters and the layer height to represent our data.
\end{abstract}

\maketitle
Granular matter is a model material presenting the phenomenology of a wide class of complex fluids with a yield stress governing the transition between solid-like and liquid-like behaviors. In this respect, the most important challenge is to establish microscopically founded constitutive relations describing the different phases as well as the features of phase transition itself. In the last decade, various important propositions were made to draw analogies with glassy states of matter or out of equilibrium thermodynamics \cite{LN98,MK02} but a full understanding still remains an open question. An important reason has been the failure of classical rheometers, which do not lead to homogeneous flows~\cite{HC05}. For dense granular flows, the situation has progressed by the introduction of new geometries (\cite{GDR04} and references therein) like the inclined plane configuration, in which the stress tensor is controlled. Now, it is known that sufficiently far from the jamming transition, the dense granular flow rheology is {\it local} in first approximation: it is governed by an effective friction coefficient increasing with the shearing rate, properly rescaled by the local confining pressure. The kinetic theory, valid in the gaseous regime,  fails to describe this situation characterized by the existence of long term contacts and force networks. In particular, binary collisions lead to a decrease of friction \cite{L01} instead of the observed increase. The most important open issue concerns the jamming/unjamming transition and the lack of identification of parameters controlling this hysteretic transition \cite{DD99}. It has recently been suggested that the dynamics in the metastable region could be dominated by elementary rearrangements, interacting non-locally \cite{KD03} by fluctuations or by the coupling with the modes of vibration\cite{DFPPPP05} (from internal or external origin).

Here we have designed a specific experiment to measure the influence of controlled vibrations on the granular rheology. We hereafter study the unjamming induced by horizontal shaking due to the transfer of energy by friction. Note that the majority of previous experiments have used vertical vibration to unjam and agitate granular matter but relatively few have used horizontal shear \cite{T01,SP01,MJN00,RM02}, see also \cite{AT06} for review.  We focus on two different questions. How does a granular film spread under the action of horizontal vibrations? How can the rheology be deduced from the spreading dynamics?
\begin{figure}[t!]
\includegraphics{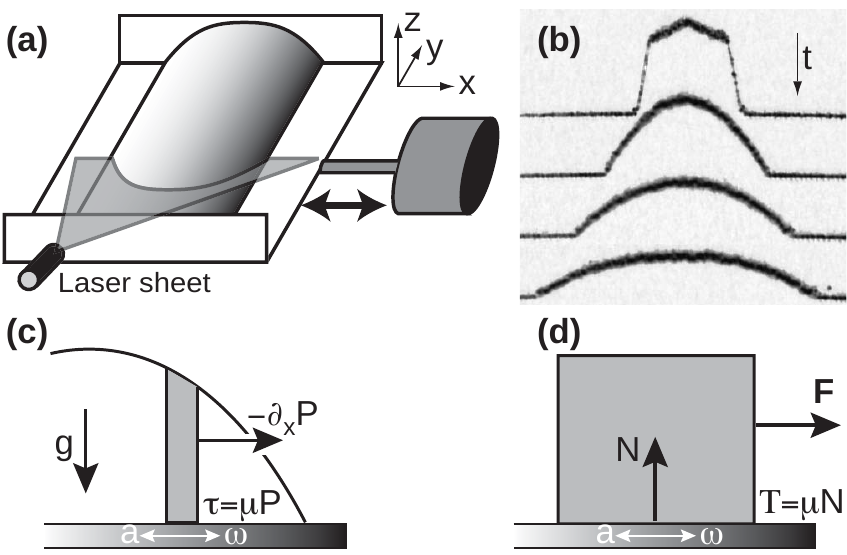}
\caption{(a) Sketch of the experimental setup. (b) Laser profile as a function of time: the top one corresponds to the granular film prior to shaking, the three below are subsequent deposit shapes after the vibration onset. Due to the laser inclination, the height is magnified by a factor $\simeq 10$. (c) Forces acting on a vertical slice of the drop. (d) Analogy with a solid slider on a vibrating tray.}
\label{setup}
\end{figure}

{\it Setup~--} The experimental apparatus is an horizontally-shaken tray (Fig.~\ref{setup}a) whose motion is a sinusoid of amplitude $a$ and angular frequency $\omega=2 \pi f$. Driving frequencies $f$ were between $15$~Hz and $30$~Hz and the maximum amplitude that we could reach is $1$~mm. The substrate is a sand-cast aluminum plate with roughness of about $10~\mu$m. The tray is stabilized by 4 ball-bearing rings gliding on two rails. The tray is leveled horizontally within $1/100^\circ$. The CCD video camera is moving along with the tray. The granular material is $d \approx 300~\mu$m diameter faceted Fontainebleau sand of static and dynamical friction coefficients $\mu_s=0.66$ and $\mu_d=0.60$. The following protocol is used to prepare a layer with reproducible initial conditions: an initial mass of sand is poured in a rectangular bottomless confining box fixed to the substrate. The box inner base defines the initial width $L_x=40$ mm in the vibration direction $x$ and $L_y=150$ mm in the lateral dimension. To level off the deposit, the granular material is shaken vigorously for $1$ min at $40$ Hz. After removal of the confining box, an initial roof-like shape sand layer is obtained (Fig.~\ref{setup}b, top). To keep a lateral confinement of the deposit, two small guiding sidewalls are fixed on the tray. The shape of the deposit is monitored by a laser sheet shined at a small angle in the central part of the layer. On Fig.~\ref{setup}b, we display below the initial profile, three subsequent laser slices observed for ``vigorous'' shaking. One sees that the granular layer looses its stability and consequently, spreads horizontally along the $x$ direction.

{\it An inertial tribometer~--} From direct visualization of the layer motion using a fast camera ($500$~Hz), we have noticed that within an oscillation period, all the surface grains bear almost the same relative phase with respect to the plate. This suggests a close analogy with the motion of a solid slider. If a constant Coulomb static friction $\mu_{s}$ is assumed, the onset of relative motion between the substrate and the solid would correspond to a rescaled acceleration $\Gamma=a\omega ^{2}/g$ equals $\mu_s$. Actually, the situation turns out to be more complex. We leave a systematic study of the motion onset to a future report and we focus here on the high $\Gamma$ regime for which the film is set into motion and eventually leaves the limit of the tray before stoppage.
\begin{figure}[t!]
\includegraphics{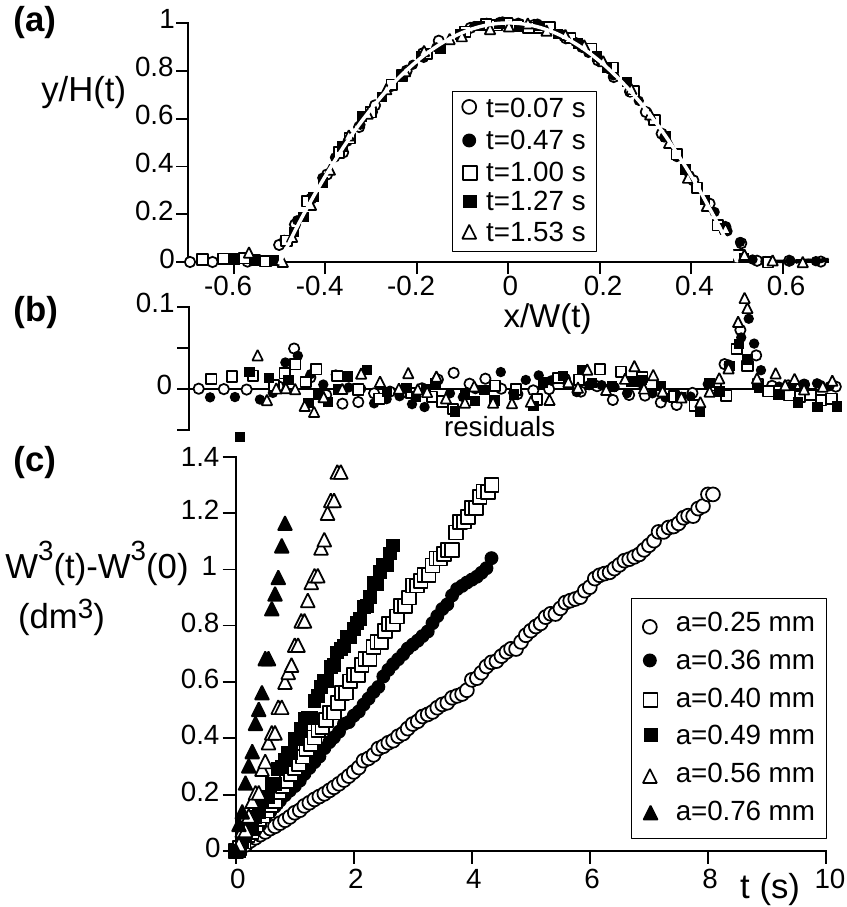}
\caption{Spreading dynamics at a frequency $f=28$~Hz for a section $S=90\pm4~{\rm mm}^2$. (a) Rescaled droplet shapes at different times $t$ for vibration amplitudes $a=0.25$~mm and $a=0.39$~mm. (b) Fit residuals. (c) $W(t)^{3}-W(0)^{3}$ as function of time $t$ for various vibration amplitudes $a$. }
\label{parabola}
\end{figure}

Let us consider the dynamics of a vertical slice of the granular droplet of width $dx$, of length $L_y$ and of height $h(x,t)$ (Fig.~\ref{setup}c). Along the vertical direction, pressure $P$ balances gravity $g$ and reads: $P=\rho g (h-z)$, where $\rho$ is the density of the material. Assuming normal stress isotropy, the pressure gradient induces a driving force $- L_y dx \rho g h \partial_x h$ on the slice. The tray exert on it a resistive force $L_y dx \tau$ ($\tau$ is the shear component of stress)  that opposes, on the average, the spreading motion. To assess the momentum transfer due to the complex sliding dynamics between the droplet and the base, we introduce a friction coefficient $\mu$ relating $\tau$ to $P$. It is worth noting that $\mu$ can depend on the basal pressure $P$, on $h$, on $a$ and $\omega$. The strong hypothesis is that it weakly depends on the relative velocity between the grains and the tray. We checked for instance that the introduction of a hysteresis between static and dynamic regimes does not affect significantly the results in the range of $\Gamma$ investigated here.

\begin{figure*}[t!]
\includegraphics{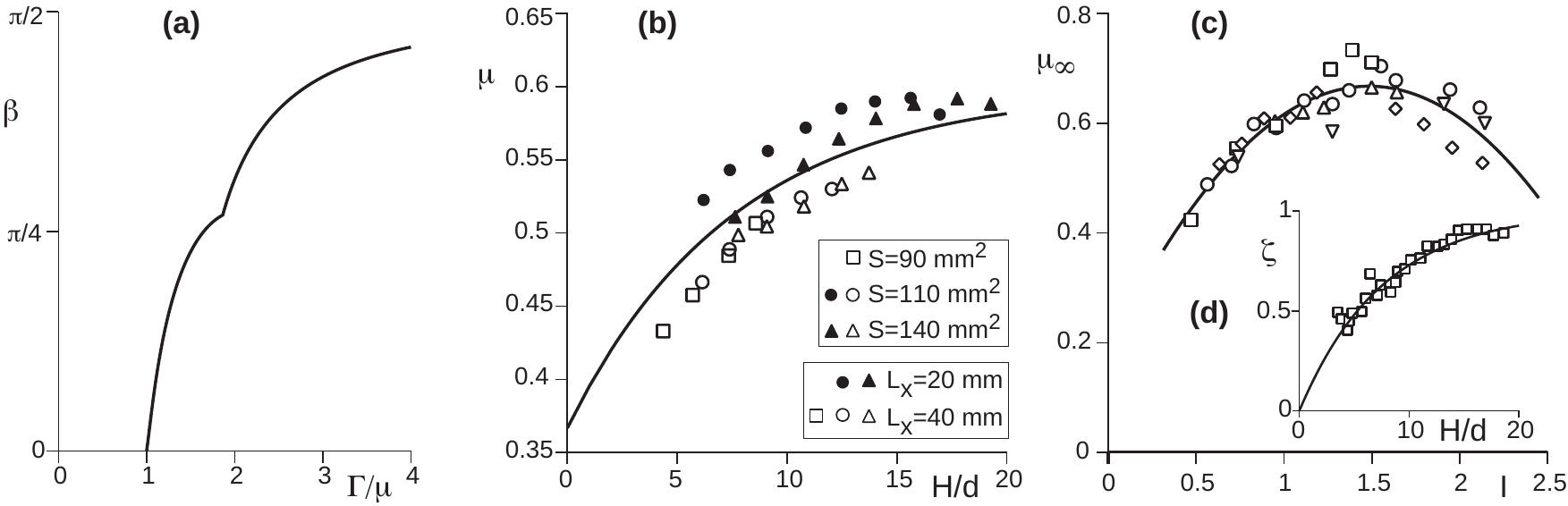}
\caption{(a) Function $\beta(\Gamma/\mu)$ relating the sliding velocity to friction (see Eqs.~(\ref{beta}),(\ref{high}) and (\ref{low})). It allows to measure an effective basal friction $\mu$ from the spreading dynamics. (b) Dependence of  $\mu$ on the drop height $H$  for $f = 24~$Hz, $\Gamma = 0.93$ and different values of $L_x$ and $S$. The solid line is the best fit using eq.~(\ref{mulaw}). (c) Effective friction $\mu_{\infty}$ in the limit of a large height $H$ as a function the rescaled vibration velocity $I=a\omega/\sqrt{gd}$ for $f=18$~Hz ($\triangle$), $f=22$~Hz ($\triangledown$, $f=24$~Hz ($\circ$), $f=26$~Hz ($\diamond$) and $f=28$~Hz ($\square$). 
The solid line is the best fit by the parabola $\mu_P(I)=0.67-0.22\,(I-1.5)^2$. (d) Height dependence of the effective friction extracted from the parabolic fit of all data: $\zeta(H/d)=\left( \mu_P -\mu\right) /\left( \mu_P -\mu_0\right)$. Each square is an average over $50$ experiments. The solid line is the best fit by the form $\zeta(x)=1 - \exp \left(-x/\mathcal{N}\right)$, with $\mathcal{N}\simeq 7.8$.}
\label{mueff}
\end{figure*}

Then, the problem is completely similar to that of a solid block on an oscillating tray, submitted to a driving force $F$ and a normal force $N=mg$ (Fig.~\ref{setup}d). The analogy is established through the dimensionless parameter  $F/N=-\partial_x h$. Our setup is in some sense the equivalent of the `inertial tribometer' \cite{baumberger} designed in the context of solid on solid friction studies. We make the equations of motion dimensionless using $\omega^{-1}$ as a characteristic time and $a$ as a characteristic length. In the sliding regime, the equation governing the evolution of the dimensionless velocity difference $v$ between the slider and the tray is:
\begin{equation}
 \Gamma \dot v=\sin t - \mu \frac{v}{|v|}+ \frac{F}{N}
\label{dynmodel}
\end{equation}
Blockage of motion occurs for $v=0$. From rest (in the moving reference frame), motion starts when: $\left| \sin t \right| =\mu/\Gamma$. We have obtained exact solutions of the problem at the linear order in $F/N$. For the purpose of the present Letter, we only need the average sliding velocity $\bar v$, which is written under the form:
\begin{equation}
\bar v=\frac{\beta[\Gamma/\mu]}{\mu} \frac{F}{N}
\label{beta}
\end{equation}
Simple calculations show that depending on the value of $\Gamma/\mu$, the block motion may undergo two different dynamical regimes: a continuously sliding at large acceleration, for which
\begin{equation}
\beta= \frac{\pi}{2} \sqrt{1-\frac{\pi^2 \mu^2}{4\Gamma^2} }
\label{high}
\end{equation}
and stop/start motions at smaller  acceleration, for which
\begin{equation}
\beta = \frac{\mu (\phi-\arcsin (\mu/\Gamma))^2 }{2 \pi \Gamma}
\label{low}
\end{equation}
where $\phi$ is the root of the transcendental equation: $\cos (\phi) -\sqrt{1-(\mu/\Gamma)^2} +\mu/\Gamma(\phi-\arcsin(\mu/\Gamma))=0$. As shown on Fig.~\ref{mueff}(a), $\beta$ is an increasing function of $\Gamma/\mu$ that reflects the fraction of time during which the solid block slides.

{\it Spreading dynamics~--} A small  aspect ratio $H/W$ droplet allows for a depth integrated  (Saint-Venant) description. Assuming no significant variation of density $\rho$, the flux of grains across a vertical section of the drop can be approximated by $a\omega \bar v h$. Using Eq.~(\ref{beta}), the mass conservation law yields a nonlinear diffusion equation:
\begin{equation}
\partial_{t}h= \partial _x (U\partial_x h^{2})\quad {\rm with}\quad U=\frac{a\omega \beta(\Gamma/\mu)}{2\mu}
\label{diff}
\end{equation}
Note that the diffusion coefficient $U$ is velocity independent. Assuming that  $U$ does not depend on height $h$, Eq.~(\ref{diff}) admits an exact parabolic self-similar solution:
\begin{equation}
h(x,t)=\frac{3 S}{2 W(t)}\left[ 1-\left(\dfrac{2x}{W(t)}\right) ^{2}\right]
\label{scale}
\end{equation}
with a spreading dynamics that can be written:
\begin{equation}
W(t)^3=W(0)^3 +72 S U t
\label{width}
\end{equation}
It can be shown that sufficiently localized initial conditions converges at long time toward the self-similar solution --~just like the convergence toward gaussians in linear diffusion. Experimentally, the profile shape is extracted from images of the laser trace, using a standard correlation technique.  We have fitted it by a second order polynomial in order to determine the width $W(t)$ and the maximal height $H(t)$. Fig.~\ref{parabola} shows the profiles measured at different times, rescaled by $W$ and $H$. Remarkably, all experimental points collapse on the predicted parabolic shape (Eq.~\ref{scale}). Residuals of the difference between the fitted curve and the experimental data shows a deviation to a parabola of less than $4\%$ in the central part and around $10\%$ on the edges, for all data presented here. As expected from mass conservation, the cross section area remains constant in time and is equal to $S=2H(t)W(t)/3$.

Several curves $W(t)^{3}-W(0)^{3}$ obtained for $f=26$~Hz are displayed on Fig.~\ref{parabola}c. They present a quasi linear time dependence whose slope increases with the vibration amplitude $a$. We have chosen here initial width values $W(0)$ such as to avoid the non-universal initial transient. Note that in our  experiment, because of the lateral boundaries, we have evidenced weak but systematic transverse curvatures: the part in the center being the fastest spreading one. Strictly speaking, the spreading dynamics is three-dimensional but may be approximated  by a two-dimensional description since the curvature effect is weak. So, theoretical predictions of the shape (Eq.~\ref{scale}) and the spreading law (Eq.~\ref{width}) are both in excellent agreement with experimental findings, Fig. \ref{parabola}.

{\it Parametric study~--} Using Eqs.~(\ref{diff}) and (\ref{width}), the slope of the relation between $W^3$ and time $t$ can be written as:
\begin{equation}
\frac{d W^3}{dt} = \frac{18gS}{\pi f} \frac{\Gamma }{\mu }\beta (\Gamma /\mu )
\label{alpha}
\end{equation}
From this relation we get an effective value of the friction coefficient $\mu$. Therefore, as initially desired, this spreading experiment under controlled vibrations can be used as a ``granular tribometer'' in order to monitor the effective friction of a vibrated granular layer. If we choose the grain diameter $d$ as the characteristic size and $\sqrt{d/g}$ as the characteristic time, there remains three control parameters, the amplitude $a$, the angular frequency $\omega$ and the cross section surface $S$ of the droplet and one local parameter evolving in time, $h/d$.

From the  local slope of $W(t)^3$, we have extracted a ``local'' effective friction coefficient and observed a systematic decrease of $\mu(t)$, as $H(t)$ decreases. Note that in the whole range of parameters, the coefficient $\mu$ remains within $30\%$ of the dynamical friction coefficient $\mu_d$. Although a clear dependence on $h$ is observed, the dynamics can  be thought of as ``adiabatic'', the shape remaining close to a parabola and the local slope of $W(t)^{3}$ giving the friction in the central part of the drop. Fig.~(\ref{mueff})b shows the effective friction coefficient extracted from spreading experiments at constant external driving parameters, starting from different initial shapes and different surfaces $S$. A tiny systematic dependence on $W$ may be evidenced, indicating  limit of the adiabatic assumption. One could iterate the analytical procedure in order to include explicitly the local height dependence in the spreading law but this goes beyond the scope of the present paper. Hundreds of experimental measurements  were performed that may all be described by the phenomenological formula:
\begin{equation}
\mu=\mu _{\infty }+\left( \mu_{0}-\mu _{\infty }\right) \exp \left(-H/(\mathcal{N} d)\right)
\label{mulaw}
\end{equation}
treating  the friction for very deep  layers $\mu _{\infty}$ as a free parameter, but with the same values for $\mu_{0}\simeq 0.37$ and $\mathcal{N}\simeq 7.8$. Figure ~\ref{mueff}c shows the fit coefficient  $\mu _{\infty}$ as a function of the dimensionless parameter that collapses the best our data, $I=\frac{a\omega}{\sqrt{gd}}$. The physical implications of this result is discussed later on. Using a parabolic approximation of  $\mu_\infty(I)$, we extract for all the experiments the thickness dependence of the effective friction (fig.~\ref{mueff}d). We are finally able to propose an empirical friction law:
\begin{equation}
\mu=\mu _{0}+\left( \mu_P\left( I\right) -\mu _{0}\right) \zeta (h/d)
\label{mulaw2}
\end{equation}
that provides fair representation of our spreading experiments.

{\it Interpretation~--} We have designed an experiment that is, to our knowledge, the first explicit measurement of a granular rheology under controlled vibration. Similarly to sheared dense granular flows, the rheology depends on an inertia number $I$ built as the ratio of a shear velocity at the scale of the grain (here $a\omega$) and the typical impact velocity. However, this last parameter was found to scale on $\sqrt{P/\rho}$ in dense granular flows \cite{GDR04}. Here, the presence of maximum in $\mu _{\infty}(I)$ allows to dismiss this possibility. We find the impact velocity to scale on $\sqrt{gd}$ (a free fall from a grain size) instead of $\sqrt{gh}$. Provided this surprising difference, we observe a regime at low inertial number where the effective friction increases with $I$, as sheared dense flows\cite{GDR04}. Moreover, at larger $I$, a cross-over to a weakening regime is evidenced. Such an effect could be interpreted as a transition from a dense regime ('liquidÕ phase) to a kinetic regime ('gaseousÕ phase). Indeed, the generic prediction of kinetic theory is a decrease of the friction with the agitation (and thus $I$) \cite{L01}. The rheological form of Eq.~(\ref{mulaw2}), featured on Fig.~\ref{mueff}c, has recently been obtained from a mean field theory in \cite{A07}. In this model, the liquid regime is dominated by the trapping of grains and the gaseous regime by binary collisions, so that the collision rate increases at low $I$ and decreases at large $I$. Another puzzling point is the finite size dependence when the layer gets thin. This nonlocal effect, involving about $\mathcal{N}=8$ grains is reminiscent of the rheology of thin sheared layers but here, goes qualitatively in the other direction: thinner layers bear less basal friction. In further studies, it will be crucial to use this `inertial tribometer' to investigate in details vibrated granular matter changing the nature and the shape of the grains and also getting closer to the jamming transition. 

We thank A.~Roussel and F.~Naudin for introducing us to the industrial problem and the Tarkett-Sommer Company for financial help. I.A. was supported by US DOE, Office of Science, contracts DE-AC02-06CH11357. I.S. was supported by PCP Franco-Venezuelan contract ``Dynamics of granular materials''.



\begin{thebibliography}{99}
\bibitem{LN98} A.Liu, S.R.Nagel, Nature {\bf 396}, 21 (1998).
\bibitem{MK02} H.A. Makse and J. Kurchan, Nature {\bf 415}, 614 (2002).
\bibitem{HC05} N. Huang et al., Phys. Rev. Lett.{\bf94}, 028301 (2005).
\bibitem{GDR04} G.D.R. Midi, Eur. Phys. Jour. E {\bf 14}, 341 (2004).
\bibitem{L01} M.Y. Louge, Phys. Fluids {\bf 13}, 1213 (2001).
\bibitem{DD99} A. Daerr and S. Douady, Nature {\bf 399}, 241 (1999).
\bibitem{KD03} A. Kabla and G. Debregeas Phys. Rev. Lett., {\bf 90}, 258303 (2003).
\bibitem{DFPPPP05} F. Dalton et al., Phys. Rev. Lett. {\bf 95}, 138001 (2005).
\bibitem{T01} G. Metcalfe et al. Phys. Rev. E {\bf 65}, 031302 (2001).
\bibitem{SP01} C. Saluena and T. Poeschel, Eur. Phys. Jour. E {\bf 1}, 55 (2001).
\bibitem{MJN00} M. Medved, H. M. Jaeger, S. R. Nagel, Europhys. Lett. {\bf 52}, 66 (2000).
\bibitem{RM02} P. Reis and T.Mullin, Phys. Rev. Lett. {\bf 89}, 244301 (2002).
\bibitem{AT06} I.S. Aranson and L.S. Tsimring, \rmp {\bf 78} (2006)
\bibitem{baumberger} T.Baumberger  et al. Rev. Scient. Inst {\bf 69}, 2416 (1998).
\bibitem{A07} B. Andreotti, submitted to Eur. Phys. Lett. (2007).
\end{thebibliography}
\end{document}